\documentclass[conference,final]{IEEEtran}
\usepackage{amsmath, amsthm}
\usepackage{color}
\usepackage[normalem]{ulem}
\usepackage{anyfontsize}
\usepackage{times}
\usepackage[latin1]{inputenc}
\usepackage{amsfonts}
\usepackage{psfrag}
\usepackage{amsbsy}
\usepackage{amssymb}
\usepackage[mathscr]{eucal}
\usepackage{latexsym}
\usepackage[T1]{fontenc}
\usepackage[dvips]{graphicx}
\usepackage{bm}
\usepackage{soul}
\usepackage{array}
\usepackage[linesnumbered,lined, algoruled]{algorithm2e}
\usepackage{showlabels}
\usepackage{dsfont}
\usepackage{tikz}
\usepackage{mfirstuc}
\DeclareMathAlphabet{\mathpzc}{OT1}{pzc}{m}{it}

\usepackage{verbatim,mathrsfs, cite}
\usepackage[acronym,shortcuts]{glossaries}

\usepackage{setspace}
\usepackage{balance}
\usepackage{showlabels}

\makeglossaries

\newacronym{ACK}{ACK}{acknowledge}
\newacronym{ARQ}{ARQ}{automatic repeat request}
\newacronym{AWGN}{AWGN}{additive white Gaussian noise}
\newacronym{BS}{BS}{base station}
\newacronym{BCC}{BCC}{broadcast channel with confidential messages}
\newacronym{BBU}{BBU}{base band unit}
\newacronym{CDF}{CDF}{cumulative distribution function}
\newacronym{C-RAN}{C-RAN}{cloud radio access network}
\newacronym{CSI}{CSI}{channel state information}
\newacronym{CSI-FB}{CSI-FB}{\ac{CSI} feedback}
\newacronym{CSIT}{CSIT}{channel state information at the transmitter}
\newacronym{DEC}{DEC}{decoder}
\newacronym{ENC}{ENC}{encoder}
\newacronym{GMP}{GMP}{Gaussian message-passing}
\newacronym{RGMP}{RGMP}{randomized GMP}

\newacronym{HARQ}{HARQ}{hybrid automatic repeat request}
\newacronym{iid}{i.i.d.}{independent and identically distributed}
\newacronym{IR-HARQ}{IR-HARQ}{incremental redundancy HARQ}
\newacronym{LDPC}{LDPC}{low density parity check}
\newacronym{MIMO}{MIMO}{multiple input multiple output}
\newacronym{MI}{MI}{mutual information}
\newacronym{MRC}{MRC}{maximal ratio combining}
\newacronym{MMSE}{MMSE}{minimum mean square error}
\newacronym{MP}{MP}{message passing}
\newacronym{MT}{MT}{mobile terminal}
\newacronym{NACK}{NACK}{not acknowledge}
\newacronym{PDF}{PDF}{probability density function}
\newacronym{PLS}{PLS}{physical layer security}
\newacronym{PMF}{PMF}{probability mass function}
\newacronym{RRH}{RRH}{radio remote head}
\newacronym{RTD-HARQ}{RTD-HARQ}{repetition time diversity HARQ}
\newacronym{SHARQ}{SHARQ}{secure HARQ}
\newacronym{S-HARQ}{S-HARQ}{secure HARQ}
\newacronym{S-QCF}{S-QCF}{secure quantized CSI feedback}
\newacronym{SNR}{SNR}{signal to noise ratio}

\theoremstyle{definition}

\usepackage[user]{zref}

\newcounter{revc}
\makeatletter \zref@newprop{revcontent}{} \zref@addprop{main}{revcontent}
\zref@newprop{revsec}{} \zref@addprop{main}{revsec}
\newcommand{\revi}[2]{%
\zref@setcurrent{revsec}{\thesection}%
\zref@setcurrent{revcontent}{#2}%
\refstepcounter{revc}%
\label{#1}
\zlabel{#1}%
\uline{#2} }
\newcommand{\revinu}[2]{%
\zref@setcurrent{revsec}{\thesection}%
\zref@setcurrent{revcontent}{#2}%
\refstepcounter{revc}%
\label{#1}
\zlabel{#1}%
#2 }

\newcommand{\revr}[2]{%
\zref@setcurrent{revsec}{\thesection}%
\zref@setcurrent{revcontent}{#2}%
\refstepcounter{revc}%
\zlabel{#1}%
\label{#1} \sot{#2}} \makeatother

\begin{document}
\sloppy

\title{Centralized and Distributed Sparsification for Low-Complexity Message Passing Algorithm in C-RAN Architectures}

\author{
	\IEEEauthorblockN{Alessandro Brighente and Stefano Tomasin}
	\IEEEauthorblockA{Department of Information Engineering, University of Padova  \\
                             via Gradenigo 6/B, 35131 Padova, Italy. \\ Email: alessandro.brighente@studenti.unipd.it, tomasin@dei.unipd.it }}

\maketitle

\begin{abstract}
\Ac{C-RAN} is a promising technology for fifth-generation (5G) cellular systems. However the burden imposed by the huge amount of data to be collected (in the uplink) from the \acp{RRH} and processed at the \ac{BBU} poses serious challenges. In order to reduce the computation effort of \ac{MMSE} receiver at the \ac{BBU} the Gaussian \ac{MP} together with a suitable sparsification of the channel matrix can be used. In this paper we propose two sets of solutions, either centralized or distributed ones. In the centralized solutions, we propose different approaches to sparsify the channel matrix, in order to reduce the complexity of \ac{MP}. However these approaches still require that all signals reaching the \ac{RRH} are conveyed to the \ac{BBU}, therefore the communication requirements among the backbone network  devices are unaltered. In the decentralized solutions instead we aim at reducing both the complexity of \ac{MP} at the \ac{BBU} and the requirements on the \acp{RRH}-\ac{BBU} communication links by pre-processing the signals at the \ac{RRH} and convey a reduced set of signals to the \ac{BBU}.
\end{abstract}

\glsresetall


\glsresetall

\section{Introduction} 

The fifth-generation (5G) of mobile communication systems has ambitious targets in terms (among others) of data rate, latency, number of supported users. Among the technologies envisioned to this end,  \ac{C-RAN} may provide the flexibility in the deployment and planning of the network, combined with powerful energy-efficient computational resources \cite{CRAN}. 

Indeed, since the signal processing of multiple cells is implemented in the centralized facility of the \ac{BBU}, the computational resources are allocated on demand to the areas that have instantaneously  more users, also with a better handling of inference and hand-off capabilities. On the other hand the need to process signals of many \acp{RRH} poses significant challenges to the \ac{BBU}. Various approaches have been proposed to reduce the huge amount of data that is exchanged in this centralized approach, including suitable quantization of either the received signal \cite{Tomasin} or the log-likelihood ratios (LLRs) \cite{Myamoto15}. On the other hand, also the signal processing itself at the \ac{BBU} is very challenging, since even a \ac{MMSE} receiver requires the inversion of very large matrices. Similar problems are encountered in massive-\ac{MIMO} systems with a huge number of users. About the reduction of signal processing burden in up-link detection, it has been proposed  in \cite{Fan} to cluster both users and \acp{RRH} based on the distance of terminals from \ac{RRH} thus parallelizing \ac{MMSE} operations into small size matrix operations. A further step forward has been done in \cite{gmp} where it is proposed to implement the \ac{MMSE} receiver by the \ac{MP}. By exploiting the Gaussian distribution of the noise, a simple solution is obtained where the complexity per unit network area remains constant with growing network sizes. In particular \cite{gmp} combines \ac{MP} with the sparsification approach of \cite{Fan}, i.e., a first selection of users based on their distance from \ac{RRH} reduces the size of the equivalent channel matrix before \ac{MP} is applied.

In this paper we leverage on the results of \cite{gmp} to propose two sets of solutions, either centralized or distributed ones. In the centralized solutions, we propose different approaches to sparsify the channel matrix, in order to reduce the complexity of \ac{MP}. However these approaches still require that all signals reaching the \ac{RRH} are conveyed to the \ac{BBU}, therefore the communication requirements among the backbone network  devices are unaltered. In the decentralized solutions instead we aim at reducing both the complexity of \ac{MP} at the \ac{BBU} and the requirements on the \acp{RRH}-\ac{BBU} communication links by pre-processing the signals at the \ac{RRH} and conveying a reduced set of signals to the \ac{BBU}.

The rest of the paper is organized as follows. We first introduce the system model in Section II. Then we propose the centralized sparsification techniques in Section III. The decentralized sparsification methods are discussed in Section IV. Numerical results are presented in Section V, before conclusions are obtained in Section VI.

Notation: matrices and vectors are denoted in boldface. $\bm{x}^T$ and $\bm{x}^H$ denote the transpose and Hermitian of vector $\bm{x}$, respectively.

\section{System Model}

We consider the up-link of a cellular network with $N_c$ cells, each one containing a \ac{BS} equipped with $N_a$ omnidirectional receive antennas (\acp{RRH}). Each cell is populated by $N_u$ \acp{MT} uniformly distributed over the entire cell area, each one equipped with a single antenna and transmitting with power $P$. 

The overall network can be seen as a \ac{MIMO} system, where the unit-power column vector $\bm{x}$ of size $K=N_c N_u$ comprises the data signals of \acp{MT} scaled by $\sqrt{P}$ before transmission, whereas column vector $\bm{y}$ of size $N = N_c N_a$ comprises all signals received by \acp{RRH}. The \ac{MIMO} channel model of the up-link from \acp{MT} to the \acp{RRH} can be written as
\begin{equation}
\bm{y}=\sqrt{P}\bm{H}\bm{x}+\bm{w}\,,
\label{recSig}
\end{equation}
where $\bm{H}$ is the $N \times K$ channel matrix with entries $[\bm{H}]_{i,j}$ and $\bm{w}$ is the \ac{AWGN} vector with \ac{iid} complex Gaussian entries with zero-mean and variance $N_0$. 

The signals received by the \acp{RRH} are forwarded to the \ac{BBU} that aims at performing the \ac{MMSE} receiver, i.e., computing 
\begin{equation}
\hat{\bm{x}} = P^{\frac{1}{2}}\bm{H}^{H}(P\bm{H}\bm{H}^{H}+N_{0}\bm{I})^{-1}\bm{y}.
\label{MMSE}
\end{equation}

\subsection{Randomized Gaussian MP decoder}

The  \ac{MP} algorithm can be used to solve the interference problem over sparse factor graphs \cite{Kschischang}, therefore providing the solution of the \ac{MMSE} receiver (\ref{MMSE}). Since the received signal is affected by Gaussian noise we can use the \ac{GMP} solution, and in particular we focus on the randomized \ac{RGMP} of \cite{gmp} which has been shown to have better convergence properties. 
In order to obtain the \ac{MMSE} estimate of the transmitted signal $\bm{x}$ the proposed \ac{RGMP} Algorithm exploits the knowledge of the statistical description of all the elements in (\ref{recSig}) and iteratively updates the values of mean and variance of all components of both $\bm{x}$ and $\bm{y}$ vectors.
The Algorithm stops updating these values when a stopping criterion is satisfied and the \ac{MMSE} estimate of $\bm{x}$ is returned.

The computational complexity of the \ac{RGMP} Algorithm is $\mathcal{O}(NK^2)$, hence it  depends on the number of users (growing quadratically with it) and receiving antennas of the system. In large systems, with many \acp{MT} and \acp{RRH}, the decoding process is therefore prohibitively complex. An approach to reduce the complexity is to reduce the number of non-zero entries in $\bm{H}$ over which the \ac{MP} is run, i.e. applying \ac{MP} on a sparsified version of $\bm{H}$. Note that the sparsification on the one side will reduce the complexity, while on the other side provides an approximation of $\hat{\bm{x}}$, thus reducing the ASR (ASR) of the system. 

Different approaches will be analysed in the following sections: a centralised approach, where sparsification is performed at the \ac{BBU} pool before \ac{RGMP} decoding, and a distributed approach, where sparsification is applied as pre-coding operations at each \ac{BS}.

\section{Centralized Sparsification Methods}
With centralized sparsification methods the decoding process is entirely demanded to the central \ac{BBU} pool. Then the signal received at the \ac{RRH}, down-converted to base-band and converted to the digital domain, is entirely forwarded to the \ac{BBU}. Hence no local processing is performed at the \ac{BS}.
Since no pre-processing operation is done at the \ac{BS} in order to reduce the computational complexity of the decoding process, this latter task is demanded to the central \ac{BBU}. We here introduce and discuss different approaches to sparsify the channel matrix by performing operations on its entries at the \ac{BBU}. 

\subsection{Sparsification based on the received power (CRPS)}

The first approach is based on the received power. In particular, we set to zero the channel matrix coefficients having power below a threshold value $P_{\min}$.

We thus obtain matrix $\hat{\bm{H}}$ with entries 
\begin{equation}
[\hat{\bm{H}}]_{i,j} = \begin{cases}
[\bm{H}]_{i,j} & \mbox{if $|[\bm{H}]_{i,j}|^2 \geq P_{\min}$} \\
0 & \mbox{otherwise} .
\end{cases}
\end{equation}

The neglected coefficients can be accounted for as additional noise into the system. In particular, defining the error matrix $\tilde{\bm{H}}=\bm{H}-\hat{\bm{H}}$ the statistical power of noise and error $N_0$ becomes
\begin{equation}
\hat{N_0}= N_0+\frac{1}{N}\sum_{n=1}^{N}\sum_{k=1}^{K}|[\tilde{\bm{H}}]_{n,k}|^2
\label{eq:n0mod}
\end{equation}

 \ac{RGMP} is then run over channel $\hat{\bm{H}}$ and considers as noise power $\hat{N_0}$.

\subsection{Sparsification based on semi-orthogonality (MCOS)}

The second proposed approach is based on \ac{MT} channels semi-orthogonality. Let us consider singularly each \ac{BS}: we notice that \acp{MT} having orthogonal channels do not interfere. Now, assuming that each \ac{MT} signal is mainly detected by the antennas of its cell, we can ignore the contribution of the external \acp{MT} since they will not significantly contribute to the computation of the \ac{MMSE}. 

In formulas, let us consider the channel row vector $\bm{h}_{k_1}=[\bm{H}]_{n_1,k_1},[\bm{H}]_{n_2,k_1},...,[\bm{H}]_{n_{N_a},k_1}]$  from \ac{MT} $k_1$ to all \acp{RRH} belonging to a certain BS with indexes in the set $\mathcal{A}=\{n_1,n_2,...,n_{N_a}\}$. The orthogonality among channels toward the same \ac{BS} is established by the internal product of the channels and we consider that two channels are semi-orthogonal if the product is below a threshold $T_{\rm prod}$, i.e.,
\begin{equation}
|\bm{h}_{k_1}  \bm{h}_{k_2}^H|^2< T_{\rm prod}.
\label{eq:tprod}
\end{equation} 

If \acp{MT} $k_1$ outside the cell $i$ is semi-orthogonal to all \acp{MT} inside the cell, then entries of channel matrix $\bm{H}$ corresponding to the link between \ac{MT} $k_1$ and all \acp{RRH} of \ac{BS} $i$ are set to zero.

\subsection{Sparsification based on the correlation}

The idea is to reduce the number of rows of the channel matrix by selecting the subset $\mathcal{S}$ of the antennas $\mathcal{A}(c)$ located in cell $c$. In order to chose a suitable subset and, hence, which rows to delete, we exploit the algorithms presented in \cite{Molish}, i.e. correlation based sparsification (CBS) and mutual information based sparsification (MIBS). We denote by $N_{ac}$ the number of antennas, and hence the number of rows of the channel matrix relative to $c$ used for decoding.

In formulas, we consider couples $\{n_1,n_2\}$ of antennas and channel matrix rows $\bm{g}_{n \in \mathcal{A}(c)}=[[\bm{H}]_{n,1},[\bm{H}]_{n,2},...,[\bm{H}]_{n,K}]$, belonging to set $\mathcal{A}$ of cell $c$ and measure their correlation as
\begin{equation}
c_{n_1,n_2} = |\bm{g}_{n_1} \bm{g}_{n_2}^H|^2.
\label{eq:corr}
\end{equation}
 For each cell the correlation between couples of antenna channels belonging to the considered cell is computed. Then the couple with highest correlation is selected and the antenna channel with lowest power is discarded. Its corresponding row in the channel matrix is hence set to zero. This procedure is repeated until we set to zero a number of rows equal to $N_a-N_{ac}$. A description of this method is provided in Algorithm 1.

\begin{algorithm}[htbp]
 \KwData{$\bm{H}, N_{ac}$}
 \KwResult{$\bm{H}$ }
 
 \For{c=1 to $N_c$}{
 \For{n=1 to $N_a-N_{ac}$}{
 \begin{enumerate}
  \item compute correlation for each couple $\{n_1,n_2\} \newline \in \mathcal{A}(c)$ with (\ref{eq:corr}), 
  \item choose the couple with highest correlation, 
  \item set to zero the row of $\bm{H}$ corresponding to the \newline antenna $n = \underset{n \in \{n_1,n_2\}}{\mathrm{argmin}}\sum_{k=1}^{K}|[\bm{H}]_{n,k}|2$
  \end{enumerate}
  }
 }\
 
\caption{Correlation Based Method (CBS)}
\end{algorithm}

\subsection{Sparsification based on the mutual information}

This antenna selection approach, MIBS, behaves similarly to Algorithm 1, except that correlation in step 1 is substituted by the normalized mutual information. The mutual information for a couple $\{n_1,n_2\} \in \mathcal{A}(c)$ is computed as
\begin{equation}
I(n_1,n_2) = \log_2 \bigg( \frac{ \| \bm{h}_{n_1}  \| ^ 2 \| \bm{h}_{n_2}  \| ^ 2}{ \| \bm{h}_{n_1}  \| ^ 2 \| \bm{h}_{n_2}  \| ^ 2 + | \bm{h}_{n_1}\bm{h}_{n_2}^H |^2} \bigg)
\end{equation}
whereas its normalized version is 
\begin{equation}
I_0(n_1,n_2) = \frac{I(n_1,n_2)}{\min \{|\log_2\left \|\bm{h}_{n_1}\right \|^2|, |\log_2\left \|\bm{h}_{n_2}\right \|^2| \}}.
\label{eq:MIBS}
\end{equation}
In Algorithm 1 we replace (\ref{eq:corr}) with (\ref{eq:MIBS}).  In both CBS and MIBS the noise power $N_0$ is not modified as in (\ref{eq:n0mod}), because, when deleting an antenna channel (and hence a channel matrix row), we assume that its information is contained in the other rows of the considered couple.

\section{Distributed Sparsification Methods}
The centralized sparsification approach has the drawback that the entire received signal is forwarded from \acp{RRH} to the central \ac{BBU}. Since the requirements for a front-haul link are very stringent (multi-gigabit-per-second-capacity and few-milliseconds latency \cite{Bartlet}) and this amount of data turns out to be prohibitively high for satisfying this requirements, we consider distributed sparsification solutions, which aim together at reducing both the decoding computational complexity and the amount of data flowing through the front-haul. 
\newline In this section we will discuss sparsification applied as pre-coding at the \ac{BS} of each cell before forwarding the received signals to the  \ac{BBU}. Let $\bm{y}_{c}$ be the received $N_a$-size column vector signal at the BS of cell $c$. If we consider a pre-coding $N_r \times N_a$ matrix $\bm{B}$ for cell $c$ and we multiply it by the received signal we obtain
\begin{equation}
\bm{B}\bm{y}_{c} = \bm{B}\bm{H}_{N_r}\sqrt{P}\tilde{\bm{x}}+\bm{B}\bm{H}_{\bar{N_r}}\sqrt{P}\bm{i}+\bm{B}\bm{w},
\label{eq:prec}
\end{equation}
where $\tilde{\bm{x}}$ is the vector containing signals coming from \acp{MT} in set $\mathcal{M}$ (as later discussed), $\bm{H}_{N_r}$ is the sub-channel matrix composed by the columns of $\bm{H}$ for users considered in $\mathcal{M}$, $\bm{H}_{\bar{N_r}}$ is the sub-channel matrix composed by the column of $\bm{H}$ for users $\not \in \mathcal{M}$ and is the vector containing signals coming from users $\not \in \mathcal{M}$.

Pre-coding matrix $\bm{B}$ can assume different forms and consider different number and types of users. In particular, we let $\bm{G}$ be the sub-channel matrix of users in $\mathcal{M}$. Then we set $\bm{B} = \bm{G}^H$, i.e. $\bm{B}$ assumes to form of the matched matrix to the considered channel. A second option provides that $\bm{B}$ is the zero-forcing matrix, i.e. $\bm{B}=\bm{G}^H(\bm{G}\bm{G}^H)^{-1}$. In the following we define different strategies to select $\mathcal{M}$.

\subsection{Selection based on the position (PSS)}

We first assume the knowledge of users location and, in particular, we know the cell each user belongs to. Then $\mathcal{M}$ is the set of users located in cell $c$, with $|\mathcal{M}|=N_u$. Matrix $\bm{B}$ will hence be a $N_u \times N_a$ dimesnion matrix. Such a pre-coding operation hence reduces the number of rows of the sub-channel matrix of each cell from $N_a$ (the number of antennas of the considered \ac{BS}) to $N_u$. We notice that, with the pre-coding operation, noise vector entries are correlated and that the \ac{MP} algorithm must be modified.  Since noise power remains the same in all branches the noise level depends on $n$ and becomes 
\begin{equation}
N_0 (n) = N_0  \sum_{k=1}^{N_u}|[\bm{B}]_{n,k}|^2,
\label{eq:n0n}
\end{equation}
with $n \in \{1,...,N_u\}$, which takes into account correlation introduced by matrix  $\bm{B}$ in each receiver branch. This new version of RGMP will be considered as default for henceforth presented methods. Note that this approach is sub-optimal respect to MMSE as the \ac{MP} solution in this case neglects the correlation among the noise components.

\subsection{Selection based on received power (DRPS)}

In this approach \acp{MT} are selected according to the received power. We select the $N_p$ users with highest power reaching the BS of cell $c$, i.e. given the channel from user $k$ to the BS in $c$, we compute the received power (\ref{eq:Rec_pow}) for each user in the cellular network, 
\begin{equation}
p(k) = \sum_{n \in \mathcal{A}(c)}|[\bm{H}]_{n,k}|^2
\label{eq:Rec_pow}
\end{equation}
and consider the $N_p$ users with highest $p(k)$ toward the \ac{BS} of cell $c$. The channel matrix columns of this set of users will then compose the columns of matrix $\bm{G}$ for cell $c$.

\subsection{Selection based on mixed criterion (MSS)}

The third approach is a mix of the first two. In fact matrix $\bm{G}$ collects columns of both users located in cell $c$ and the $N_p$ most powerful users, i.e. with highest $p(k)$, located outside cell $c$.

\section{Numerical Results}

We here first present the ASR results obtained for all the sparsification methods introduced in previous sections and then discuss their computational complexity. Mostly the trade-off between ASR and computational complexity is analyzed. We consider a scenario with $N_c=16$ cells, each one equipped with a BS with $N_a = 8$ \acp{RRH}. Each cell contains $N_u=4$ users and each user is allocated the same transmitting power $P=1$. Noise power is chosen to have a border cell \ac{SNR} of $0$ dB. In the following we assume that $\bm{H}$ is affected by both path loss (with  coefficient $\alpha=2$) and Rayleigh fading, so that each entry is a zero-mean complex Gaussian random variable with variance equal to the inverse of the distance from the considered \ac{MT} and the considered antenna of the \ac{BS}. Channel matrix entries are i.i.d.

The \ac{RGMP} Algorithm is stopped when the mean of the transmitted signal does not change more than 1\% in one iteration.  Each method has been compared both in terms of sparsification level, i.e. the number of entries of the channel matrix $\bm{H}$ different from zero after sparsification, and channel ASR. All results have been compared with those of pure \ac{RGMP}, i.e. without channel sparsification. 

\subsection{Centralized sparsification}

\begin{figure}[htbp]
\centering
\includegraphics[width=1\hsize]{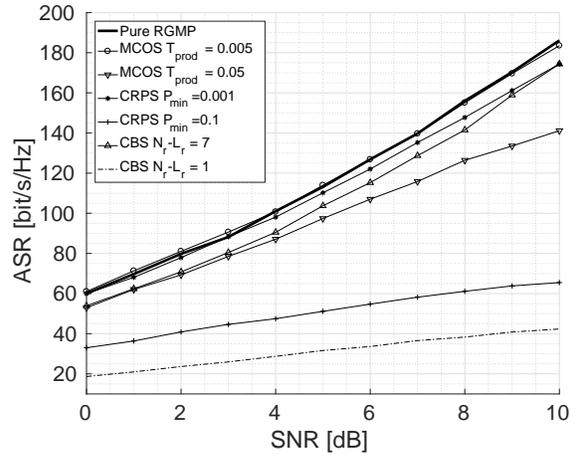}
        \caption{Mean ASR vs. \ac{SNR} for centralized sparsification.}
        \label{fig:cent}
\end{figure}

We consider first the centralized sparsification. \newline Fig. \ref{fig:cent} reports the mean ASR values vs. \ac{SNR} for two parameter values of each centralized sparsification method and for \ac{RGMP} without channel sparsification. ASR results for MIBS are analogous to the ones obtained with CBS, and are not reported here for brevity. With all the presented methods we can obtain good results in terms of ASR values, comparable or equal to that obtained with \ac{RGMP} without channel sparsification.

\subsection{Distributed sparsification}
\begin{figure}[htbp]
\centering
\includegraphics[width=1\hsize]{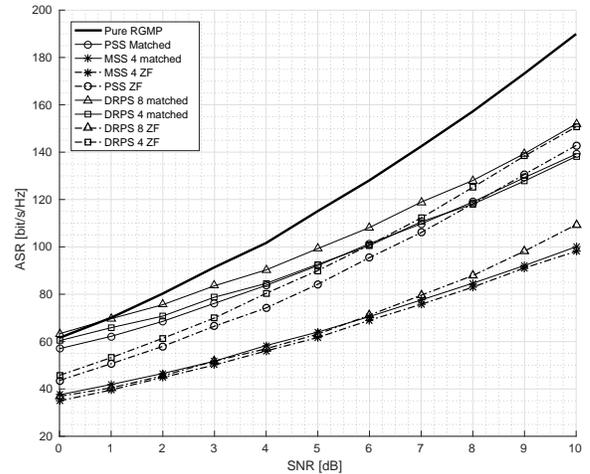}
        \caption{Mean ASR vs$SNR$ for distributed sparsification.}
        \label{fig:dist}
\end{figure}

Distributed sparsification has been implemented for both matched and zero forcing matrix $B$. Fig. \ref{fig:dist} reports mean ASR values vs. \ac{SNR} obtained for the maximum and minimum considered users by distributed sparsification methods and for \ac{RGMP} without channel sparsification. We denoted the different methods with their acronym followed by the number of considered users. We can see that the matched implementation of $B$ outperforms the zero-forcing implementation is terms of mean ASR. Furthermore the matched implementation of all methods considering the maximum number of users, allows a better exploitation of the channel for low \ac{SNR} values obtaining mean ASR values equal to the ones obtained with \ac{RGMP} without channel sparsification.

\begin{table*}[t]
\begin{center}
   \caption{Best decoding computational complexity and ASR for the different sparsification methods: $0$ dB \ac{SNR}}
\begin{tabular}{ | c |  c | c |  c |  c | }
  \hline	
   			
   \textit{Sparsification method}  & \textit{Sparsification level} & \textit{\# of operations} & \textit{ASR [bit/s/Hz]}  \\ \hline
   Pure RGMP  & 8192  & 2097152 & 60 \\ \hline
   CRPS, $P_{\rm min}=0.001$  & 4121  & 1582464 & 60.19 \\ \hline
   MCOS, $T_{\rm prod}=0.001$ &  4288  & 1097728 & 58.83 \\ \hline
   CBS, $L_{r}=1$ &  7168 & 1835008 & 55.72\\ \hline
   MIBS, $L_{r}=1$ &  7168 & 917504 & 58.73 \\ \hline
   PSS, $\bm{B}=\bm{G}^H$ & 4096  & 1835008 & 57\\ \hline
   MSS, 6 usr.$\bm{B}=\bm{G}^H$ & 6144  & 1835008 & 63.5\\ \hline
   DRPS , 4 usr. $\bm{B}=\bm{G}^H$ & 4096 & 1966080 & 61.28\\ \hline
   DRPS, 4 usr. $\bm{B}=\bm{G}^H(\bm{G}\bm{G}^H)^{-1}$ & 4096  & 393216 & 46.2\\ \hline
   DRPS , 8 usr. $\bm{B}=\bm{G}^H$ & 8192 & 4082131 & 63.25\\ \hline
\end{tabular}
\end{center}

  \label{tab:compComp}
\end{table*}

\subsection{Computational complexity analysis}

\begin{figure}[htbp]
\centering
\includegraphics[height = 8cm,width= 10 cm]{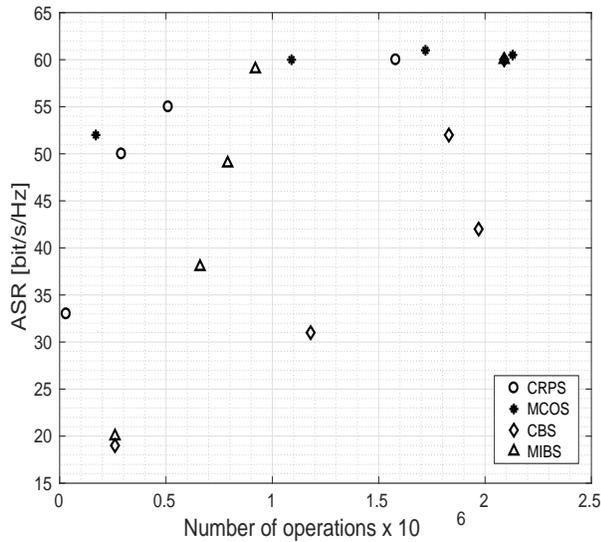}
        \caption{Trade-off between computational complexity and achievable sum-rate for centralized sparsification methods: $0$ dB \ac{SNR}.}
        \label{fig:comp0}
\end{figure}

We now analyse the computational complexity of the different approaches in terms of number of decoding operations after sparsification. This depends on the number of entries of $\hat{\bm{H}} \neq 0$ as each  requires two sums over the total number of users $K$, operations that are repeated until the stopping criterion is satisfied. Hence the total number of operations is
\begin{equation}
N_{op} = 2 \, K \,s \, I,
\end{equation}
where $s$ denotes the number of channel matrix entries different from 0, and $I$ the number of message passing iterations needed to satisfy the stopping criterion. 
Fig. \ref{fig:comp0} shows the ASR vs. the number of operations needed for the decoding process for the centralized sparsification methods with an \ac{SNR} level of 0 dB. We notice that with semi-orthogonal-based sparsification we obtain the best performing system, with an achievable sum rate of 58 bit/s/Hz and a computational complexity of $9.2 \cdot 10^5$ operations. However notice that this implementation is not the best performing in terms of achievable sum rate, instead it is the best compromise between computational complexity and ASR. Notice that \ac{RGMP} without channel sparsification obtains an ASR of 60 bit/s/Hz with a computational complexity of $2.1 \cdot 10^6$ operations. Hence the reduction of $1 \cdot 10^6$ operations comes with an ASR loss of 2 bit/s/Hz. 

\begin{figure}[htbp]
\centering
\includegraphics[height = 7cm,width= 9 cm]{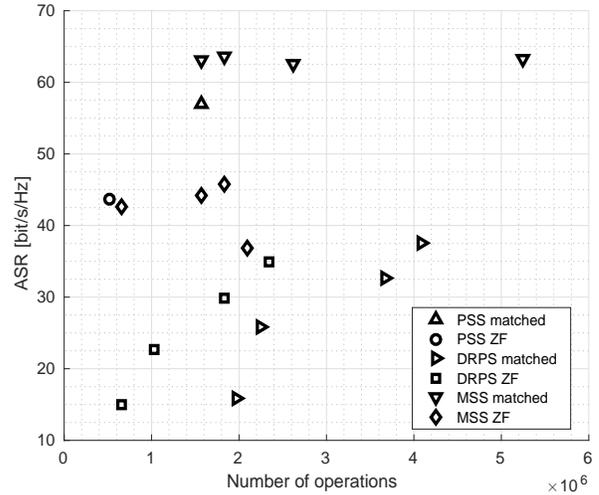}
        \caption{Trade-off between computational complexity and achievable sum-rate for distributed sparsification methods: $0$dB \ac{SNR}}
        \label{fig:compa_0}
\end{figure}

Fig. \ref{fig:compa_0} reports the ASR vs. the number of operations needed for the decoding process for the centralized sparsification methods
with an \ac{SNR} level of 0 dB. We notice that the best compromise between ASR and computational complexity is obtained for MSS with matched matrix, which presents an ASR of approximately 63 bit/s/Hz with a computational complexity of $2 \cdot 10^5$ operations. 

Table I reports the obtained computational complexity and ASR values for the best performing parameter of each method when \ac{SNR} value is $0$ dB. A trade-off can be obtained, since we want to maximize  the ASR while maintaining a low computational complexity.
We can hence state that all methods present a channel ASR comparable to the one obtained with pure \ac{RGMP}, but generally need a significantly lower number of decoding operations. The best performing among all presented methods in terms of both computational complexity and ASR is MIBS sparsification when \ac{SNR} value is $0$ dB. This method needs less than half of the number of operations required by pure \ac{RGMP} with an ASR loss of approximately 2 bit/s/Hz.

\balance

\section{Conclusions}

For a \ac{C-RAN} system where signals coming from many \acp{RRH} we have considered the problem of implementing a \ac{MMSE} receiver at the \ac{BBU}. In order to decrease the computational complexity a \ac{RGMP} algorithm has been considered, and suitable sparsifications of the channel matrix have been introduced. We considered both centralized approaches, performed at the \ac{BBU} and requiring a complete transfer of received signals from the \acp{RRH}  and decentralized solutions where a pre-processing is performed at the \ac{BS}. This latter solution not only has been shown to be effective in terms of reduction of the computational complexity of the decoding process, but also of the amount of data flowing from the \acp{BS} to the \ac{BBU}, and hence of the front-haul network capacity as well as the centralization overhead. Numerical results have shown a variety of trade-off between complexity and performance (in terms of ASR) confirming that the proposed solutions are promising for an implementation of these approaches in 5G \ac{C-RAN} systems.

%\bibliographystyle{IEEEtran}
%\bibliography{biblio}

\end{document}